\documentclass[useAMS,usenatbib,letterpaper]{mn2e}

\usepackage{graphicx}
\usepackage{natbib}
\usepackage{aas_macros}
\begin{document}

\title[Radial velocity mapping of 
AW~UMa]{Radial velocity mapping of Paczy\'nski's
star AW~UMa:\\ Not a contact binary\thanks{Based on the data 
obtained at the David Dunlap Observatory, University of Toronto}}

\author[Theodor Pribulla and 
Slavek M. Rucinski]{T. Pribulla$^{1,2}$\thanks{E-mail: pribulla@ta3.sk (TP)}
and S.M. Rucinski$^{2}$\thanks{E-mail: rucinski@astro.utoronto.ca (SMR)} \\
$^1$Astronomical Institute of the Slovak Academy of Sciences, 059\,60 Tatransk\'{a} 
Lomnica, The Slovak Republic\\
$^2$Department of Astronomy, University of Toronto, 
50 St.~George St., Toronto, Ontario, Canada M5S~3H4
}

\date{Accepted 0000 Month 00. Received 0000 Month 00; in original form 2007 March 17}

\pagerange{\pageref{firstpage}--\pageref{lastpage}} \pubyear{2008}

\maketitle

\label{firstpage}

\begin{abstract}
We present two-dimensional (radial velocity, orbital phase) 
spectroscopic results for the very low mass-ratio 
close binary AW~UMa which strongly indicate that the 
spectroscopic mass ratio ($q_{sp} = 0.10$) 
does not agree with the photometrically derived one and that
the widely adopted contact binary model appears to
experience serious inconsistencies and
limitations for this object. AW~UMa is
compared with V566~Oph ($q_{sp} = 0.26$) which we found to behave
according to the contact model. Observed broadening 
functions of AW~UMa can be interpreted by a very strong
limb darkening and/or non-solid body rotation of the
dominant primary component; the former assumption is
unphysical while the differential rotation 
is not supported by an apparent stability of localized, 
dark features on the outer side of the primary.  
There are indications of the existence of an
equatorial belt encompassing the whole system. 
All deficiencies in the interpretation 
and the discrepancy between the photometric and spectroscopic
mass ratio of AW~UMa can be solved within a new model
of AW~UMa where both components are detached 
and the system is submerged 
in a stream of hot, optically thick matter which 
mimics the stellar contact. While the masses and their ratio
are correctly given by spectroscopy, the photometric picture
is heavily modified by the matter engulfing both stars in the
equatorial plane.

\end{abstract}


\section{INTRODUCTION}
\label{intro}

Most of the light-curve analyses of contact binary stars
of the W~UMa type assume --
following the seminal papers of \citet{Lucy1968a,Lucy1968b}
-- that shapes of these stars are well described by 
a common equipotential surface of the Roche model. However, 
light curve analyses have a common limitation: 
a light curve is a result of a mapping of a complex
three-dimensional surface into an one-dimensional function 
of time. The information content of many contact binary 
light curves is low, particularly when total eclipses
do not occur. The correctness of the assumed model is 
of a crucial importance for a proper interpretation 
of these binaries and still remains an open issue.

The exact solid-body rotation of the Roche model 
is a strong assumption of the Lucy model. 
While it is a convenient simplification 
of the problem, we have no
theoretical basis to believe that a strictly solid-body 
rotation is really present. The Sun does not rotate
rigidly and there are many indications that other 
solar type stars also rotate differentially. We have
rather vague ideas for rotation rates some hundred times
than the solar. Simple arguments based on the 
presence of horizontal temperature gradients could be 
used to argue that convection cells should break into small 
ones leading to an enhanced turbulent viscosity and thus
a solid-body rotation. 
But -- on the other hand -- the large difference in the 
component nuclear 
energy production requires an effective transport over 
common envelope which cannot happen without extensive 
horizontal motions. Also, the degree of the solid-body
rotation may depend on the mass of the secondary component;
when it is very small, its tidal influence may be too weak
to prevent the big primary from rotating differentially. 

Within the Roche model, a common equipotential surface 
has a relatively simple shape and is described by only three 
free geometrical parameters (the mass ratio, $q$, the degree of contact, 
$F$ and the inclination, $i$); this parametric description is even 
simpler than for detached binaries where sizes of components 
are unrelated. All extant solutions of contact 
binary light curves still require a proof of the strict
applicability of the Roche model described in such simple
terms.

In this paper we attempt to lift the degeneracy of 
the 3D into 1D light-curve mapping by using the 
Broadening Functions formalism \citep{Rci2002}. 
A broadening function (BF) gives the surface brightness
distribution of a stellar object in the velocity space. 
This is true if all surface points 
radiate the same spectrum, which appears to be a particularly  
valid assumption for contact binaries (see \citet{ander1983}). 
In the case of a solid body rotation, the observed radial 
velocity is proportional to the projected distance from 
the axis of rotation. When combined with the uniform surface 
brightness, a single BF gives a 1D image of system in the velocity 
space. We go one step further and analyze several
BF's by arranging and combining them in the orbital phase 
into 2D functions (radial velocity, orbital phase),
effectively the phase dependent,
2D radial-velocity maps of a binary system. 
The approach of arranging spectra in the phase domain,
when the phase coherence is an issue, 
has already been used by several researches, 
particularly to study rapid changes in cataclysmic variables.  
In the DDO studies of the binary stars, it was used mostly
as an auxiliary tool \citep{ddo10,ddo11,ddo12} 
for faint binaries where individual
BF's were poor and some phase averaging was advantageous. 
But here, it is used as the main way to 
assess correctness of the solid-body rotation 
and to analyze the well observed contact binary AW~UMa. 

The main object of our study, AW~UMa, is presented in 
Section~\ref{obj}. The observations are described in 
Section~\ref{obs} while the determination of radial 
velocities through a Gaussian and rotational
profile fitting and through direct modelling 
of the BFs is presented in Section~\ref{velocities}. 
The resulting spectroscopic elements and the absolute
parameters are presented in Section~\ref{elements}. 
Sections~\ref{mass-ratio} and \ref{deviate} 
discuss the critical issue of the applicability of the Roche model
and of the deviations from it that we have discovered.
A new model of AW~UMa is presented and discussed in
Section~\ref{newmodel}. The results are summarized in 
Section~\ref{sum}. Throughout the paper V566~Oph is discussed as a 
comparison object for AW~UMa; this contact binary was 
previously analyzed in \citet{ddo11}. 

\section{AW~UMa}
\label{obj}

After the discovery of AW~UMa by \citet{BeP1964} and after 
the important demonstration by \citet{MD1972a}
that its photometric variability 
beautifully agrees with the then recently 
developed contact model of \citet{Lucy1968a,Lucy1968b} 
-- this being true in spite of its unexpectedly small 
value of the photometrically derived mass ratio 
$q \simeq 0.07 - 0.08$ -- the binary has played a special 
role in the field of contact binaries. Its very small
photometric mass ratio was re-confirmed several times 
(for references to observational studies, see \citet{Prb1999}). 
Throughout the development of the field, all theoretical 
studies had to include this object as a crucial if 
extreme datum in all structural and evolutionary modelling 
efforts. The key point was that even at this extreme mass ratio, 
with the almost totally energetically inert secondary component, 
the surface temperature can be equalized over the 
whole common envelope as described by an equipotential
surface of the Roche model. 

The particularly good characterization of the light curve 
and the seemingly unambiguous conclusion that $q \simeq 0.07 - 0.08$ 
were a direct result of the presence of total eclipses 
strongly constraining the parameter space of the contact 
model solutions. 
While the photometric observations leave very little 
room for questioning the low value of $q$ for AW~UMa -- as long
as it really fulfils all conditions of the Lucy
model -- spectroscopic data were less easy to interpret.
This is because the spectral lines of the primary component 
are broad and show a very small and difficult to measure 
orbital variability with the semi-amplitude of the variation, 
$K_1$, which is {\it several times smaller\/} than the 
rotational broadening of its lines. The spectral
signature of the secondary component, while nowadays rather
easily detectable, is weak indicating $L_2/L_1 \sim 0.1$ in 
the visual region. As the result, the spectroscopic elements of 
AW~UMa remain virtually undetermined in spite of several
previous attempts. 

\begin{table}
\begin{scriptsize}
\caption{Properties of AW~UMa and V566~Oph. \label{tab01}}
\begin{center}
\begin{tabular}{lcc} 
\hline
HD                 & 99946       & 163611     \\
HIP                & 56106       & 87860      \\
$\pi$ [mas]        & 15.13(0.90) & 13.98(1.11)\\
$V_{max}$          & 6.83        &  7.46      \\
$V_{min}$          & 7.13        &  7.96      \\
$B-V$              & 0.33        &  0.41      \\
$M_V$              & 2.74(12)    &  3.20(16)  \\
$M_V$ (RD)         & 2.70        &  3.08      \\
Sp. type           & F0V - F1V   &  F4V       \\
Period [d]         & 0.4387258   & 0.4096546  \\
$T_0 (HJD)$        & 2\,450\,124.9954  & 2\,451\,314.0094 \\
\hline
\end{tabular}
\end{center}
\end{scriptsize}
\medskip
The values of the period and of $T_0-2\,400\,000$ were used
for calculating phases of the spectroscopic observations. 
Absolute visual magnitude, $M_V$ was determined using 
the Hipparcos parallax, while $M_V$ (RD) is
absolute magnitude estimated using calibration 
of \citet{rd1997} from observed colour and period. 
Standard errors of some parameters are given in parentheses
\end{table}

\citet{BeP1964} was not able
to detect the secondary component and determined orbital
parameters of the primary component only by
measuring positions of $H_\gamma$ and $H_\delta$ on 
photographic spectra with a (low) dispersion of
75 \AA/mm; this resulted in $V_0 = -1 \pm 2$ km~s$^{-1}$ and 
$K_1 = 28 \pm 3$ km~s$^{-1}$. \citet{Lean1981} 
used spectra of a higher dispersion (20 \AA/mm) and
applied the cross-correlation technique. He 
marginally detected the secondary component and 
gave very preliminary spectroscopic elements: 
$V_0  = -17 \pm 7$ km~s$^{-1}$,
$K_1 = 29 \pm 8$ km~s$^{-1}$ and $K_2 = 423 \pm 80$ 
km~s$^{-1}$.  

\citet{rens85} were able to measure the 
radial velocities of the primary component stacking 
Doppler profiles of selected lines. 
Their low value of $K_1 = 22.2 \pm 0.9$ km~s$^{-1}$ 
alleviated the previous problem of the 
unexpectedly large total mass ($M_1 + M_2 \sim 4 \,M_\odot$) 
which had resulted from a relatively large semi-amplitude 
$K_1$ and the (assumed) small mass ratio. Unfortunately, 
the exposure times of the spectra were very long 
(4.7 -- 11.2 \% in phase) which resulted in  
smearing of the spectral lines and a reduction of $K_1$. 
An important point stressed by \citet{rens85} was that the
spectral lines were too narrow when compared with the
Lucy model predictions. Even the hydrogen lines
were narrow, which was difficult to interpret in any
reasonable way; the authors considered problems 
with the continuum placement as a cause which could
indeed be magnified when working with photographic
spectra. This ``narrow line problem'' had been noticed by
several researchers \citep{MD1972a,ander1983} but --
because it did not affect the photocenter velocities --
it remained an unexplained curiosity.

The first high-quality CCD spectra and their analysis were 
presented by \citet{Rci1992a} who developed the new BF
technique for this star. Even with the good-quality 
spectroscopic data, 
it was impossible to reliably determine the full set of
parameters. For that reason \citet{Rci1992a} chose to fix $q$ 
at the photometric value, determining only the overall 
velocity span, $K_1+K_2$. The data appeared to 
confirm the contact model, but some latitude in the selection 
of the best mass ratio $q$ by 0.005 around 0.075
resulted in an uncomfortably large 
range in total masses, $M_1+M_2$. Finally, \cite{Prb1999} 
obtained new photographic spectra and reanalyzed all published 
radial velocities. Subtraction of line profiles between 
the two quadratures indicated  $q \approx 0.08$. 
The authors tried to explain the differences in 
previously determined systemic velocity $V_0$ by the
multiple nature of AW~UMa and the centre-of-mass 
motion.

The current study has been mostly driven by our realization 
that new high-quality spectroscopic 
data obtained at the David Dunlap Observatory (DDO) 
for AW~UMa do not confirm the low value of the mass 
ratio suggesting a larger value, closer to $q \simeq 0.10$. 
The seemingly small change in the mass ratio 
from $\simeq 0.08$ to $\simeq 0.10$ is important: 
Not only that it strongly affects
the final value of the total mass of the system, $M_1+M_2$,
by controlling $M_1$,
but also -- as recently shown by \citet{BeP2007} --
it is practically impossible to obtain a stable structural 
model for AW~UMa for $q < 0.1$ and -- even then -- the secondary 
turns out to be a very unusual object with
a very small core and a tenuous, low density envelope.

With the mass ratio being such an important parameter, 
we decided to compare the broadening functions 
of AW~UMa with identically obtained data for the 
contact system V566~Oph. This binary is very similar 
in spectral type, orbital period and brightness to AW~UMa, 
but its mass ratio is larger, 
$q_{ph} \simeq 0.24$ \citep{MD1972b}, 
$q_{sp} = 0.26 \pm 0.01$ \citep{ddo11}.
It is also a totally eclipsing system, a condition 
which very strongly constrains and improves parametric solutions. 
The V566~Oph data utilized in the current paper were 
already presented in \citet{ddo11} and are discussed here only
when directly relevant for the discussion of AW~UMa.
We summarize the essential properties of both systems 
in Table~\ref{tab01}. 

Neither AW~UMa nor V566~Oph have close companions. We checked AW~UMa
carefully and see no very close body which would show in radial velocity 
centre-of-mass velocities. We see also no spectral signatures 
of a K- or M-type dwarf star in our spectra while unpublished 
CFHT adaptive-optics observations (see \citet{Prb2006}) 
did not reveal any close visual
companions at separations $>0.2$ arcsec. 
AW~UMa forms a common proper-motion 
pair with BD+30\degr 2164 at an angular separation of 
about 67 arcsec. The radial velocity of the 
companion, $-13.28 \pm 0.39$ km~s$^{-1}$ \citep{tokov2002}, 
is very close to the centre of mass velocity of AW~UMa. 
The proper motion velocity components for both stars are also similar: 
According to the Hipparcos Tycho project \citep{Tycho2}, 
$\mu_\alpha \cos \delta = -82.8(9)$ mas/yr and $-82.0(16)$ mas/yr 
and $\mu_\delta = -199.3(9)$ mas/year and $-198.6(15)$ mas/year,
for AW~UMa and BD+30\degr 2164, respectively. While the presence
of the distant companion of AW~UMa has no direct relevance to our
discussion, it does help in constraining the 
absolute magnitude of AW~UMa at $M_V \simeq  2.5$ (assuming 
$B-V=0.72$ for the G5 main sequence companion of AW~UMa)
which indicates the spectral type of A9V/F0V. The 
spectral type of AW~UMa is very hard to estimate directly
because of the strong broadening of the spectral lines; it is
most often quoted as F0/2V.

\section{OBSERVATIONS}
\label{obs}

All discussed here observations of AW~UMa were obtained 
using the slit spectrograph in the Cassegrain focus of 1.88m 
telescope of the David Dunlap Observatory. The spectra were 
taken in a window of about 240 \AA\ around 
the Mg~I triplet (5167, 5173 and 
5184~\AA) with an effective resolving power of 
about $12,000 - 14,000$. The journal is given in 
Table~\ref{tab02}. Two different diffraction gratings 
were used, with 1800 lines/mm and 2160 lines/mm,  
with the same effective spectral resolution,  
but with a different sampling 
per pixel (0.145 \AA/pixel and 0.117 \AA/pixel). 
One-dimensional spectra were extracted 
by the usual procedures within the IRAF 
environment\footnote{IRAF is distributed by the National 
Optical Astronomy Observatories, which are operated by 
the Association of Universities for Research in Astronomy, 
Inc., under cooperative agreement with the NSF.} 
after the bias subtraction and the flat field division. 
Cosmic ray trails were removed using a program 
provided by \citet{pych2004}. Broadening functions (BFs) 
were extracted by the method described in \citet{Rci1992a} 
using, as templates, spectra of slowly rotating standard 
stars of similar spectral type (see below). 

\begin{table}
\begin{scriptsize}
\caption{Journal of spectroscopic observations of AW~UMa and V566~Oph. \label{tab02}}
\begin{center}
\begin{tabular}{llcccc}
\hline
Object & Spectrum & HJD          & Phase  & Exp. & ADU \\ 
       &          & 2\,400\,000+ &        &      &     \\ 
\hline
AW~UMA & K0022717 & 53785.9046 & 0.9222 & 902 & 48941 \\
AW~UMA & K0022718 & 53785.9165 & 0.9494 & 901 & 40397 \\
AW~UMA & K0022720 & 53785.9291 & 0.9782 & 902 & 44140 \\
AW~UMA & K0022721 & 53785.9398 & 0.0026 & 901 & 40923 \\
AW~UMA & K0022723 & 53785.9516 & 0.0295 & 902 & 39354 \\
AW~UMA & K0022730 & 53785.9788 & 0.0914 & 825 & 33179 \\
AW~UMA & K0022882 & 53789.8271 & 0.8630 & 374 & 27166 \\
AW~UMA & K0022883 & 53789.8325 & 0.8752 & 487 & 23001 \\
AW~UMA & K0022884 & 53789.8408 & 0.8940 & 901 & 15881 \\
AW~UMA & K0022938 & 53790.5965 & 0.6167 & 546 & 28549 \\
\hline
\end{tabular}
\end{center}
\end{scriptsize}
\medskip
Explanation of columns: Spectrum - original FITS 
file name; HJD - heliocentric julian date
of mid-exposure; Phase - phase of mid-exposure; 
Exp. - exposure time in seconds; 
ADU - median level of the extracted spectrum in ADU. 
The full table is available in the 
electronic form only. 
Phases correspond to optimal ephemerides in Table~\ref{tab04} 
(for AW~UMa according to the solution 
with $u_1$ = 1.00 and $u_2$ = 0.56).
\end{table}

AW~UMa was observed on 12 nights of February 18/19 to 
April 15/16, 2006 and on April 9/10, 2007 
using the 2160 lines/mm grating. In total, 109 spectra 
uniformly covering the orbital cycle of the binary
were obtained. The exposure times varied between 5.5 and 
15 minutes (0.9 -- 2.4 \% of the orbital period). 
The star HD~128167 ($\sigma$~Boo, F2V, $V \sin i = 5$ km~s$^{-1}$,
$V_r = +0.2 \pm 0.9$ km~s$^{-1}$) served as the BF template. 
As expected, the use of a different template 
(HD~222368, $\iota$~Psc, F7V) resulted in
very similar broadening functions. 

V566~Oph was observed on 6 nights between May 5/6 and June 22/23, 
2005 with the 1800 lines/mm grating. 
These data are described in \citet{ddo11}. 
Later, the system was observed using the 2160 lines/mm grating on 
4 nights between August 22/23, 2005 and September 18/19, 2005;
in total, 57 spectra were observed. 
HD~222368 (F7V, $V \sin i = 3$ km~s$^{-1}$, 
$V_r = +5.4$ km~s$^{-1}$) was used as a template. 

\begin{figure}
\includegraphics[width=84mm,clip=]{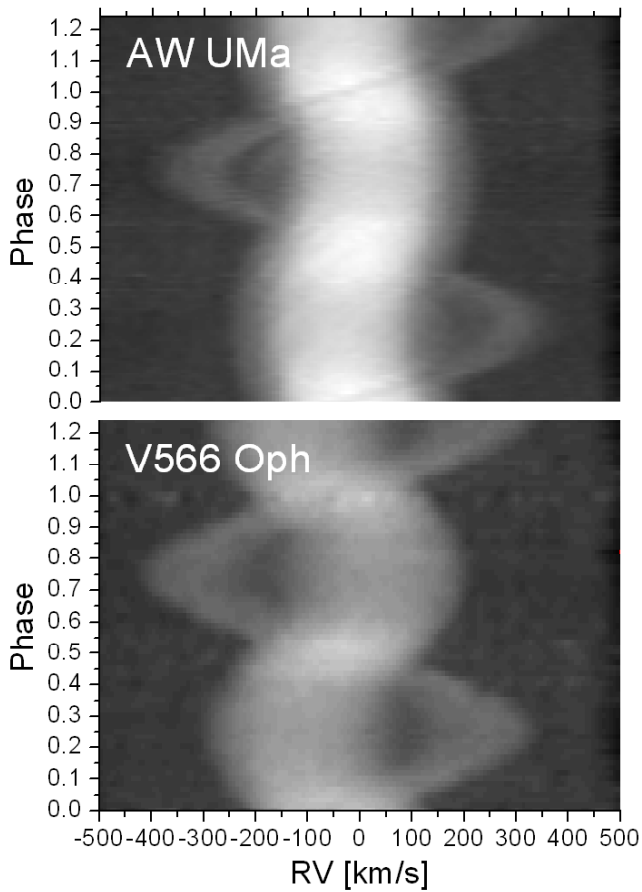}
\caption{The greyscale plot of the broadening functions of AW~UMa (top) 
and V566~Oph (bottom). The BFs were re-binned to equal 
intervals in the phase and then Gaussian smoothed 
(see the text for details). Residuals from two considered
models are shown in Figures~\ref{fig05} and \ref{fig09}.
\label{fig01}}
\end{figure}

For a direct comparison, the BFs for AW~UMa and V566~Oph
were extracted with the same step in radial velocity 
and were smoothed by convolving with the Gaussian function 
of $\sigma$ = 1.5 bin in radial velocity corresponding
to 10 km~s$^{-1}$. A 2D (radial velocity, orbital phase) map 
of AW~UMa was constructed by sorting the available 109 BFs
in phase, re-binning them with a phase step of 0.01 and, finally, 
smoothing by a convolution with the Gaussian function of
$\sigma = 0.01$ in phase; the corresponding map for
V566~Oph, because of fewer observations, was 
constructed using the phase step of 0.02.
The map for AW~UMa, in Fig.~\ref{fig01}, 
clearly shows the orbital motion of both components. 
The transit of the secondary component is seen as a dark 
streak cutting through the profile of the primary.
Here we see the first but important deviation from the
model in that the width of the secondary in the
velocity space is much narrower during the 
transit than could be expected from its appearance 
at the orbital quadratures; it is also asymmetric as
the two limb-darkened ``edges'' of the secondary are 
not equally deep. There exists also a
marked difference of the BFs between the two
quadratures which we discuss in Section~\ref{deviate}). 
We note that the phased BFs reveal presence of faint, 
localized, dark features on the primary component
drifting at the star rotation rate which are
best visible as diagonal streaks around
phases 0.3 to 0.55. We will 
speculate that these are photospheric spots which remained
stable over one year of the AW~UMa observing.

\section{RADIAL VELOCITIES AND MODELING OF THE BROADENING FUNCTIONS}
\label{velocities}


Because of the extreme mass ratio, AW~UMa is a difficult object
for radial velocity determination from the BFs, in the way described 
by \citet{Rci1992a}. All problems encountered for
contact binaries appear in this case in their most elevated form
so that their brief summary may be in order. 

There exist several ways to determine radial velocities from the
BFs. The simplest and entirely model-independent one 
is through fitting of Gaussian profiles to the peaks in
the BFs. This technique has been successfully used throughout 
the DDO series of spectroscopic element determination for 
90 binaries, see the DDO papers \citet{ddo1} to \citet{ddo10}. 
Unfortunately the Gaussian profiles do not really represent
the expected shapes of the BFs and are just a 
convenient numerical tool. Starting with the paper \citet{ddo11},
we have been using the rotational profiles. 
The theoretical rotational profiles slightly depend on the
limb darkening (which can be fixed at a reasonable value) but
-- operationally -- are very similar to Gaussians.  
Effectively, this approach corresponds to an approximation of 
the binary components by limb-darkened spheres. 
Because of the sharp, almost vertical edges of such profiles, 
they appear to give more stable and consistent centroid
results than the Gaussians. 

Both, the Gaussians and the rotational 
profiles give the central positions of the fitted
BF peaks, but neither of the approaches includes 
proximity and eclipse effects. The projected photo centres of  
binary components are close to their mass centres, but
small asymmetries (such the reflection effect, which is
however small in contact binaries or ellipsoidal
distortion asymmetries) remain and are not accounted for. 
Having the individual velocities determined, the spectroscopic orbital 
elements are determined by either fitting simple sine curves 
to the photocentric velocities  
or by modelling the BFs with the Roche model assumption leading
to predictions how proximity effects may shift the light centroids.

Instead of fitting the individual components,
the whole set of the available BFs can be simultaneously
fitted by model BFs calculated using the Roche model.
This approach is vastly more physical, but its major disadvantage is 
the large number of the parameters and complex inter-dependencies 
between them.

\begin{figure*}
\includegraphics[width=120mm,clip=]{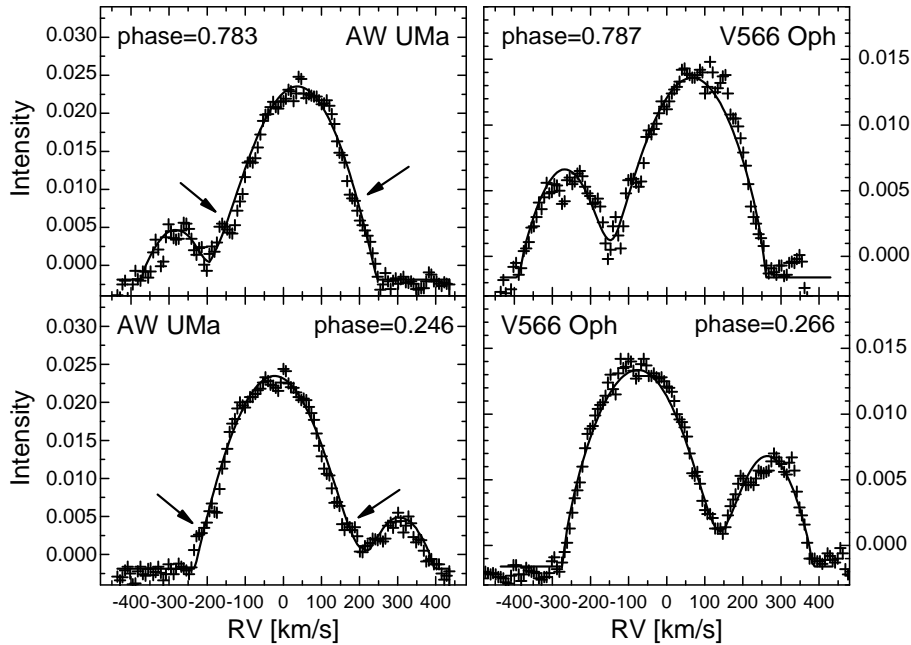}
\caption{Fitting the model profiles to typical 
broadening functions of AW~UMa and V566~Oph at  
the orbital quadratures. Note the relatively narrow and
peaked shape of the AW~UMa primary signature as well as 
the ``kinks'' which broaden it below some 30\% above the
baseline (indicated by arrows). 
The limb darkening coefficient 
was fixed at the theoretically predicted 
values, 0.52 for AW~UMa and 0.555 for V566~Oph. 
The slightly negative zero level of the broadening functions is
due to difficulties with the true continuum placement of the 
heavily broadened spectra; the {\it shapes\/} of the BFs are 
unaffected, but they do suffer vertical shifts of an unknown size.
This effect has also affected the BFs in Figures~\ref{fig07}
and \ref{fig09}. We comment on this effect in item 4 of 
Section~\ref{deviate}.
}
\label{fig02}
\end{figure*}

In the present paper we use all the above techniques to evaluate their 
effects on the spectroscopic orbit of AW~UMa. For the Gaussian and the
rotational profile fitting, we used the BFs away from the eclipses, 
beyond phases $\pm 0.12$ from the eclipse centres. For the direct 
modelling, we used all the available BFs, including those within
the eclipses. We evaluate the used techniques below in succession.

\subsection{Gaussian profile fitting}
\label{gauss}

A sum of two Gaussian functions was fitted to each BFs by adjusting
seven parameters: the baseline level, intensities of both components, 
their radial velocity positions and widths. By intensities
of components we mean their individual strengths in the BFs;
their added integrals are expected to equal unity for a
perfect spectral-type match between the object 
(AW~UMa) and the template.

\subsection{Rotational profile fitting}
\label{rotational}

In an approach similar to the Gaussian fitting, 
we added two rotational profiles to form one double-peaked 
profile; the limb darkening coefficient for both components
was set at the same value, $u_1 = u_2 = u$.
Again there are seven parameters to fit: the baseline, 
intensities of both components, their radial velocity 
positions and half-widths. Because the BF peaks for AW~UMa
had a ``pointed'' or ``triangular'' shapes, unlike
what is expected from rotation alone, we considered different
values of $u$ between $u = 0.2$ to $u= 1.0$. Unexpectedly,  
a measure of the global quality of the fit, the 
weighted sum of squares of residuals for all 
simultaneously fitted BFs, reached a minimum for 
an extreme value of $u=1$. This is entirely incompatible 
with the spectral type of F0/2V  
at the wavelength 5184 \AA\ where an appropriate value 
of the limb-darkening coefficient should be around 
0.52 \citep{hamme93}. 

It should be noted that the global fit in $u$ related
mostly to the properties of the primary profile because
the contribution of the secondary was small and its peak
was hard to fit. When left to an automatic adjustment,
the half-width of the secondary peak varied over a large range. 
We were forced to fix it at 70 km~s$^{-1}$ 
which provided globally the best fits. 
In the end, two values of limb-darkening coefficient, 
$u = 0.52$ and $u = 1.00$, were considered, the latter
being unphysical but able to reproduce the data much better
than the former.

While AW~UMa clearly shows the very strange shape of its primary
component in radial velocities (in BFs), in the
case of V566~Oph the shapes of extracted BFs are 
consistent with the rotational profiles computed
for the expected theoretical limb darkening of $u$ = 0.56. 
We show typical broadening functions at the orbital quadratures
in Fig.~\ref{fig02}. The fits shown in the figure 
are for the contact model BFs, 
as described in Section~\ref{direct}. Attention is drawn to 
the peaked shape of the AW~UMa primary and the presence
of unexplainable ``kinks'' in the profile some 30\% above
the baseline which are related to an additional broadening
at the base.

We note in passing that we did not consider any of
the non-linear forms of the limb darkening. This decision resulted
mostly from the good reproduction of the V566~Oph BFs by predictions
based on the linear limb darkening.

\subsection{Rotational profiles modified for differential 
rotation}
\label{differential}

\begin{figure}
\includegraphics[width=84mm,clip=]{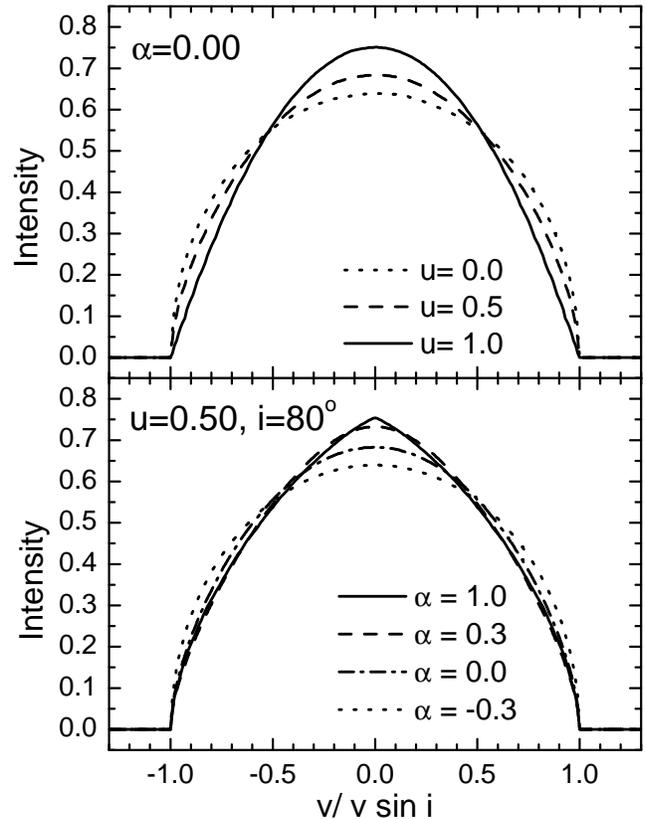}
\caption{Effects of the varying limb darkening (top) 
and differential rotation (bottom) on the shape 
of the rotational profile of a single rotating star. 
The latitude ($b$) dependence of angular velocity
was computed as $\omega (b) = \omega (0) [1 - \alpha \sin^2 b]$. 
In the case of a differential rotation, the shape 
of the rotational profile slightly depends on the 
inclination angle.
}
\label{fig03}
\end{figure}

While an artificially large value of $u$ can give a 
more peaked signature of the AW~UMa primary, an entirely
different explanation for such a shape 
can be sought in terms of the differential rotation
of the solar type, with the polar regions 
moving slower than the equatorial regions. The simplest 
assumption of a dependence of the angular rotation 
velocity $\omega$ on the latitude $b$ is:
\begin{equation}
\label{eqn01}
\omega (b) = \omega (0) [1 - \alpha \sin^2 b].
\end{equation}
For simplicity and because of the geometry of AW~UMa,
we assumed that the equatorial 
$\omega (0)$ was always equal to the orbital rotation rate.
With this assumption, $\alpha =0$ corresponds 
to the solid-body rotation while $\alpha = 1$ 
corresponds to a strongly differential rotation with 
the non-rotating poles. 
The rotational profile with such a description for 
$\omega$ cannot be expressed as an analytical function and 
requires a numerical integration over the visible 
part of star even for a simplified case of the 
star retaining the spherical shape (such as the solar case). 
The effects of the limb darkening and the
differential rotation for a spherical star
are shown in Fig.~\ref{fig03}. 
The shape of the rotational profile only
weakly depends on the inclination of the rotation axis.

\begin{figure}
\includegraphics[width=84mm,clip=]{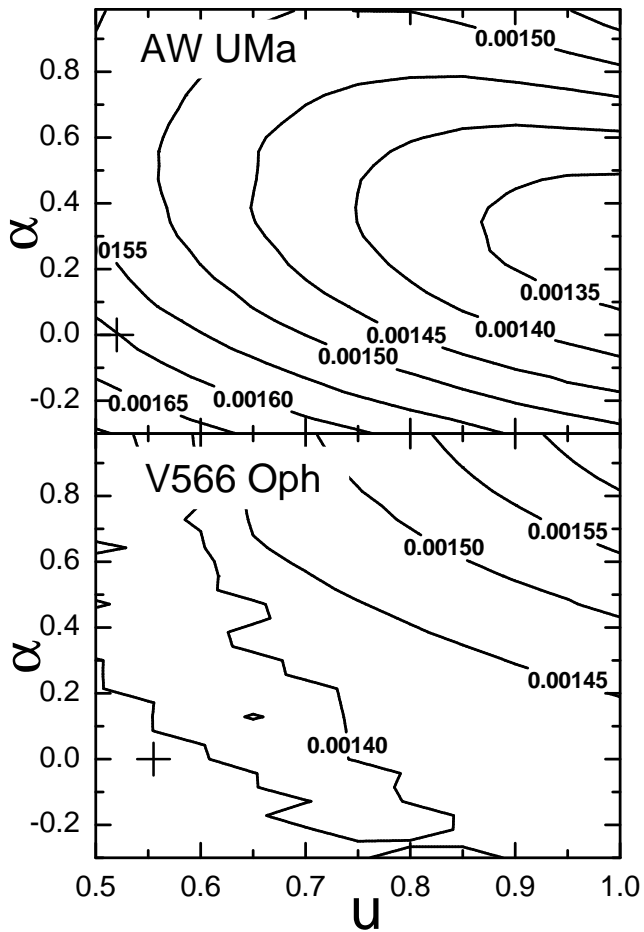}
\caption{Results of a grid search for the limb darkening 
and differential rotation ($u$ and $\alpha$) coefficients
best representing the observed BFs of AW~UMa (top). 
The optimal fits occur for the highest permissible
limb darkening coefficient of $u = 1.0$ and 
$\alpha \approx 0.3$ (the pole rotating slower than equator). 
A similar plot for V566~Oph (bottom) shows a shallow
minimum close to the nominal value of the 
limb darkening and no rotation (cross). 
}
\label{fig04}
\end{figure}

The rotational profiles with $\alpha >0$ tend to reproduce
the narrow shape of the primary in the broadening functions.
A grid search for the best values of the
limb darkening coefficient $u$ and the parameter $\alpha$ giving 
global minimum of weighted sum of squares of residuals 
for all BFs of AW~UMa outside 
eclipses was performed within the domain $0.50 < u < 1.00$ 
and $-0.3 < \alpha < 1.0$. The orbital inclination of
$i$ = 78.3$\degr$ was adopted for AW~UMa after 
\citet{Prb1999}. The resulting dependence of 
$\chi^2$ is shown in Fig.~\ref{fig04}. 
It appears that the best fit can be obtained for
an unphysical, large value of $u \rightarrow 1$ 
and for $\alpha \approx +0.3$; if a more physically 
acceptable value of $u$ within $0.5 < u < 0.6$ was 
adopted, then the fits would require an even higher degree of the
differential rotation with $\alpha \approx +0.55 \pm 0.05$. 
A similar grid search for V566~Oph in Fig.\ref{fig04} 
shows a different behaviour: The dependence of 
$\chi^2$ on both search parameters is rather weak 
and global minimum is not far from the expected
limb darkening ($u = 0.56$) and solid body rotation.

\subsection{Direct modelling of the observed BFs using
the contact model profiles}
\label{direct}

The approaches described above involved separate
measurements of radial velocities for each component, to be
later combined in a common spectroscopic orbit. One can go
one step further: By fitting synthetic BFs to 
the observed ones -- under an assumption
of the Lucy contact model -- the spectroscopic element 
determination is more direct; there is only one step of
BF fitting which incorporates orbital parameters.
In addition, this approach takes into account the 
proximity effects (such as deformation 
of components, gravity darkening and mutual irradiation) 
so that all available BFs can be
used, without a limitation to those with phases outside the
eclipses. Such an approach was used successfully before
for AW~UMa and AH~Vir by \citet{Rci1992a} and \citet{LR93}.

The modelling of the BFs for AW~UMa was performed assuming 
convective envelope for both components ($A_1 = A_2 = 0.50, 
g_1 = g_2 = 0.32$). The temperature of the primary was held 
fixed at $T_1$ = 6980~K, corresponding to spectral type F0V according 
to the calibration of \citet{pop1980}. The temperature of the secondary, 
$T_2$, was coupled to $T_1$ through the gravity
darkening law. The limb darkening coefficients were 
interpolated in the tables of \citet{hamme93} according to 
the mean surface temperature and gravity. The local surface fluxes were 
computed for the observed spectral range (5074 \AA -- 5306 \AA). The 
optimized parameters were: the mass ratio $q$, fill-out factor $F$, 
sum of semi-amplitudes of radial velocities $(K_1 + K_2)$, systemic 
velocity $V_0$ and temperature of the secondary $T_2$ (if decoupled from 
$T_1$ via gravity darkening law). We also adjusted the baseline level and 
normalization (scaling) of the BFs and the instant of the spectroscopic 
conjunction, $T_0$ (i.e., we did not use the pre-determined 
phases); the orbital period was, however, kept fixed at 
0.4387258 days. The synthetic BFs were smoothed by 
convolution with a Gaussian with $\sigma = 10.0$ 
km~s$^{-1}$ to exactly correspond to observational 
BFs (see Section~\ref{obs}). The phase smearing due to the
length of exposures was taken into account by integrating the
BFs over phase intervals corresponding to exposures with the integration step 
of 0.002 in phase. The extracted BFs of V566~Oph 
were modelled with similar assumptions, 
only $T_1$ was set to 6540~K to correspond to the F4V spectral type. 

We found that a representation of the observed 
broadening functions of AW~UMa 
by synthetic ones with the limb darkening taken from 
the tables of \citet{hamme93}, with $T_2$ coupled to $T_1$
 through the local gravity darkening, results in a surprisingly poor
match. Firstly, we experienced problems similar to that 
for the rotational profile fitting in that the primary
component showed a much more peaked signature  
than predicted. This may be taken as a proof that
the difficulties described in
Sections~\ref{rotational} and \ref{differential} were
not caused by our neglect of the proximity effects,
but were due to genuine deviations from the model.
Secondly, with the direct fitting of the whole BFs, the 
difficulties were aggravated by the additional constraint of
having the component temperatures, 
$T_2$ and $T_1$, closely linked through variations of the local
gravity. The problem can be partially (if unphysically) resolved 
by making $T_2$ a free parameter and/or by assuming unrealistically 
high values of the limb darkening coefficient and/or
of the differential rotation. Each of these ``fixes'' 
creates its own difficulties. 

\begin{figure}
\includegraphics[width=84mm,clip=]{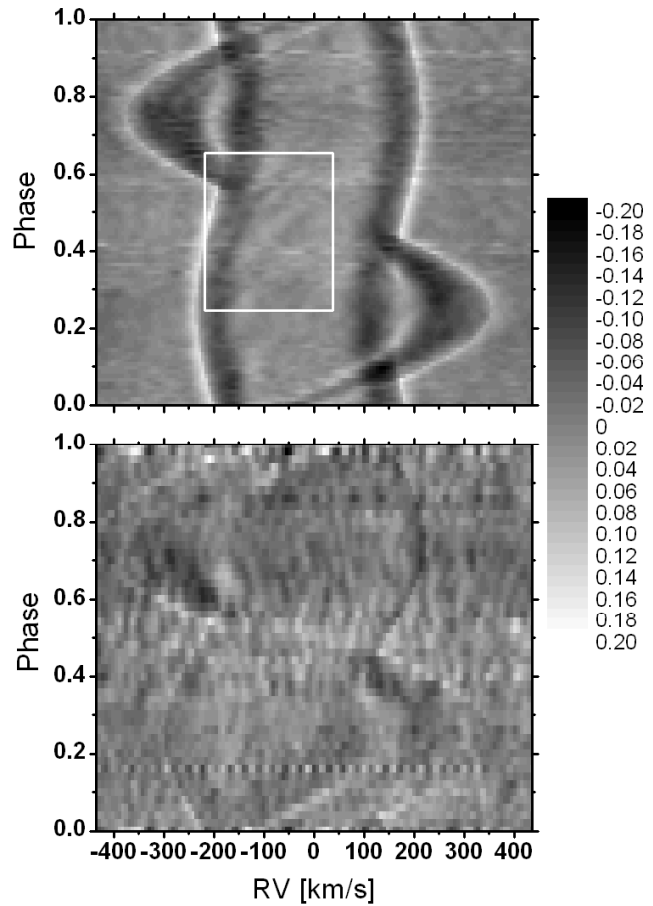}
\caption{The map of residuals from the best 
fit to the observed BFs of AW~UMa (top) and V566~Oph (bottom) 
using the direct BF fitting. For a comparison of both stars,
the residuals are shown as deviations normalized to the 
respective maxima of the BFs so that the gray scale applies
to both panels. The ``kinks'' in the AW~UMa profiles,
described in the
text and marked in Figure~\ref{fig02}, are located at the
place where the negative (dark) deviations suddenly change into
positive (bright) ones.
Although the secondary signature is always fainter than
predicted, its transit in front of the primary (phase 0.0)
does not appear as a dark trough, so that the secondary
appears to have a similar temperature to that of the
primary and a negligible limb darkening
(however, we do see an asymmetry in the
secondary transit, as commented in item 3 of
Section~\ref{deviate}).
Note the diagonally drifting dark features
in the AW~UMa residuals (in the white box) which we interpret
as photospheric spots. The most conspicuous
spot  emerges at the very left edge of the profile so it must be 
located close to the equator of the primary component. The 
spots had to be stable over the one year of the 
full span in our observations to produce such a consistent 
picture. The residuals for V566~Oph do not show
any systematic substantial deviations except a perturbation
around phase 0.65.
}
\label{fig05}
\end{figure}

To fit the observed BFs by the model ones, we used several 
combinations of assumptions, but discuss here only three:
\begin{description}
\item[(i)] A normal, i.e., expected atmospheric 
limb darkening, no differential rotation and the temperatures
$T_2$ and $T_1$ coupled through the gravity-darkening;
\item[(ii)] A large value of the limb darkening coefficient 
and $T_2$ decoupled from $T_1$;
\item[(iii)] A normal limb darkening, differential rotation
and $T_2$ decoupled from $T_1$.
\end{description} 
The last assumption requires an explanation:
The differential rotation was modelled in a simple way
by assuming the component shapes and the brightness
distribution as predicted by the Roche model, and by
modifying the rotational velocity according to the 
assumed, latitude-dependent, differential rotation law.
Although this approach is not strictly consistent
(the Roche model requires solid body rotation) it is
somewhat modelled on the Sun which remains spherical
in the presence of the $\simeq 30$\% differential rotation.
We used $\alpha$ fixed at 
several values with the best fit for AW~UMa
obtained for $\alpha = 0.30$. 
The re-binned and smoothed residuals for the case (i)
for both, AW~UMa and V566~Oph, are shown in Fig.~\ref{fig05};
the large systematic deviations for AW~UMa should be noted,
particularly in comparison with the almost ideal agreement
with the contact model for V566~Oph.
The final parameters of AW~UMa for all considered cases are 
given in Table~\ref{tab04}. 

An inspection of the residuals of fits for AW~UMa 
obtained under the above assumptions shows that the best 
representation is obtained for the case (ii) with
$u_1 = 1.00$ and with $T_2$ decoupled from $T_1$ and thus
individually optimized. Although this assumption is not 
physical, it does reproduce the observed shape of 
primary component signature the best. 
The secondary component always negligibly contributes to the
overall value of weighted sum of residuals hence its limb darkening 
could be always kept at the theoretically predicted value. 
It is interesting to note that for V566~Oph, the
BFs can be modelled without any artificial assumptions (see 
Section~\ref{deviate} for discussion). The resulting 
spectroscopic elements of V566~Oph are close to those
found from the same spectra in \citet{ddo11} by
using the rotational profile fits. While systemic
velocity is practically the same, the sum of the
semi-amplitudes, $K_1 + K_2$ is only about 1.6\% 
larger in the present, direct Roche-model modelling.

\section{Spectroscopic elements and the absolute parameters of AW~UMa}
\label{elements}

In order to determine the full range of uncertainties
related to various methods of processing the
radial velocity information,
the spectroscopic elements for AW~UMa 
were obtained first by simple sine-curve fits to individual 
radial velocities (Sections~\ref{gauss}, \ref{rotational},
\ref{differential}), then by direct fits to the whole BF
profiles (Section~\ref{direct}). 
The results of the former are given
in Table~\ref{tab03}. The results of the 
direct BF fits are tabulated in Table~\ref{tab04},
together with those for V566~Oph given for
comparison. 

\begin{table*}
\begin{scriptsize}
\caption{Spectroscopic elements of AW~UMa for three cases of 
the limb darkening and differential rotation (see text). Standard errors of 
parameters are given in parentheses. \label{tab03}}
\begin{center}
\begin{tabular}{lcccc} 
\hline
Parameter & Gaussian & Rotational & Rotational & Rotational \\
Case      &   (1)    &    (2)     &     (3)    &    (4)     \\
\hline
$u$                        &     --               & 0.52            & 1.00            & 0.52            \\
$\alpha$                   &     --               & 0.0             & 0.0             & 0.60            \\
$V_0$ [km~s$^{-1}$]        &  $-9.17(0.77)$       & $-8.64(0.53)$   & $-8.83(0.61)$   & $-8.76(0.54)$   \\
$K_1$ [km~s$^{-1}$]        &  28.45(0.92)         &  29.16(0.76)    & 29.81(0.77)     & 30.24(0.68)     \\
$K_2$ [km~s$^{-1}$]        & 312.77(2.62)         &  299.77(1.03)   & 295.06(1.71)    & 295.5(1.68)     \\
$K_1 + K_2$ [km~s$^{-1}$]  & 341.22(2.78)         &  328.93(1.28)   & 324.87(1.87)    & 325.74(1.81)    \\
$M_1 [M_\odot]$            & 1.655(40)            &  1.474(16)      & 1.415(24)       & 1.425(24)       \\
$M_2 [M_\odot]$            & 0.151(6)             &  0.143(5)       & 0.143(5)        & 0.146(4)        \\
$M_{12} [M_\odot]$         & 1.806(40)            &  1.617(17)      & 1.558(25)       & 1.571(24)       \\
$q = M_2/M_1$              & 0.091(3)             & 0.097(3)        & 0.101(3)        & 0.102(2)        \\
$\sum w(O-C)^2$            & 1774.07              & 1454.3          & 1622.3          & 1483.3          \\
\hline
\end{tabular}
\end{center}
\end{scriptsize}
\end{table*}

\begin{figure}
\includegraphics[width=84mm,clip=]{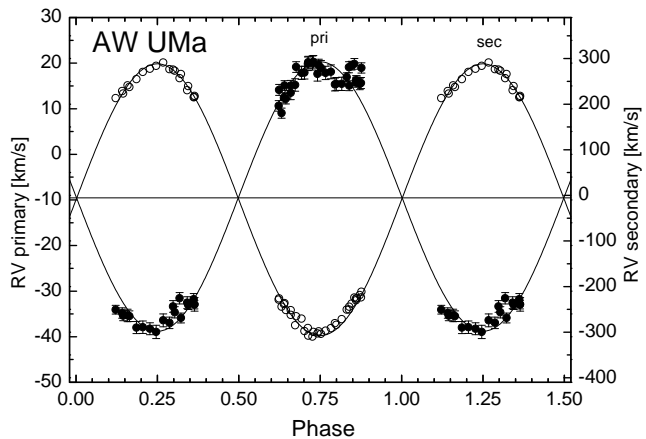}
\caption{The radial velocity curves for the
primary (full circles and the left vertical axis) 
and the secondary component (open circles and the right
vertical axis) for AW~UMa. 
The errors of radial velocities are shown for the
primary component only; the formal 
errors for the secondary component are comparable
with the size of symbols.
}
\label{fig06}
\end{figure}

The sine-curve fits were determined for the following cases:
(1)~the Gaussian fits, as described in Section~\ref{gauss}, 
(2)~the rotational profile fits with the expected limb darkening, 
$u = 0.52$ as described in Section~\ref{rotational}, 
(3)~the same, but with an artificially 
enhanced limb darkening, $u = 1.00$,
and (4)~rotational profile fits with $u = 0.52$, with 
a strong differential rotation characterized by $\alpha = 0.55$ 
(found by a grid search; see Fig.~\ref{fig04}), as described in 
Section~\ref{differential}. The elements for the four cases are 
given as columns in Table~\ref{tab03}. The phase diagram of radial 
velocities for the case (3), best representing
the observed BFs, is shown in Fig.~\ref{fig06}.

A comparison of the Gaussian and rotational profile fits 
for BFs extracted around the orbital quadratures shows that 
the rotational profiles better represent the data. Moreover, 
Gaussian fits tend to overestimate the sum of semi-amplitudes, 
$K_1 + K_2$, and the total mass (see Table~\ref{tab03}); this is
caused by a ``cross-talk'' in the sense that the wide wings of
either of the components contribute to the profile 
of the other component. Especially, the secondary component position
is driven away from the mass centre. Hence, we regard the spectroscopic
elements derived by the Gaussian fits as biased.  
The rotational profile fits give markedly 
more consistent results with the final 
spectroscopic mass ratio, $q \approx 0.10$. 

The contact model BF fits did not provide individual velocities of the
components so that the orbital elements of the binary
system were evaluated directly. We again considered various
combinations of the limb darkening coefficients (the expected
ones as well as the enhanced one, $u=1.0$, for the primary),
as tabulated in Table~\ref{tab04}. 
Again, the best representation of the observations was obtained 
with the strong limb darkening of the primary component.
Even if the large limb darkening is not a correct assumption,
it does give the best fits to the observed BFs. In the other two
cases, $\chi^2$ is dominated by differences in the 
predicted and observed shape of the primary; 
the least-squares fits ignore the shape
of the secondary component and 
its semi-amplitude is incorrectly determined.

We summarize the results in Tables~\ref{tab03} and \ref{tab04}
by noting that while sum of semi-amplitudes 
$K_1 + K_2 \sim$ 325 km~s$^{-1}$, 
which leads to the total mass of the system
of $M_1+M_2 \simeq 1.5 - 1.7 \,M_\odot$
is consistent with values found by \citet{Rci1992a}, 
the spectroscopic mass ratio, $q \sim 0.1$, appears to be always
larger than in the previous photometric investigations. 

The absolute parameters of AW~UMa, by necessity pertaining
mostly to the primary component, suggest a moderate-mass
star of an advanced evolutionary state.
The primary component mass, for the case of the
high limb darkening, $M_1 = 1.636 \pm 0.005  M_\odot$ (the formal
mean standard error),
is compatible with the Terminal Age Main Sequence position. The 
combined absolute magnitude of AW~UMa, $M_V$ = 2.63 is compatible 
with the Hipparcos parallax and the distance estimated from
the proper-motion companion (Table~\ref{tab01}). We return
to the matter of the absolute parameters of AW~UMa
in the last Section~\ref{sum}.

\section{The difference of the spectroscopic and photometric
mass ratio}
\label{mass-ratio}

In all light curve solutions for contact binaries, 
determination of the photometric mass ratio is based on an
assumption that both components are described by a  
common envelope corresponding to the same equipotential 
value within the Roche model. 
Geometrical relations for such a structure are
entirely predictable and well known. 
\citet{MD1972a} pointed out that only two parameters, 
the mass ratio -- entering via the ratio of radii -- 
and the orbital inclination uniquely determine 
the times of the inner eclipse contacts for totally
eclipsing contact systems. This beautiful idea has worked 
numerous times for many contact binaries and has 
tremendously improved quality of light curve 
solutions. As the currently ongoing spectroscopic 
program at the David Dunlap Observatory \citep{ddo12} 
can attest, contact binaries which do not show 
total eclipses frequently have light curve solutions 
giving entirely erroneous and useless values of 
$q_{ph}$ when compared with the spectroscopic, direct 
determinations of $q_{sp}$. Thus, there is no reason
to question such total-eclipse solutions as long as the
main assumption of the Lucy model of the 
common equipotential 
in the {\it strict Roche model\/} is fulfilled. 
However, the common equipotential is still an 
assumption; if the secondary underfills its
Roche lobe, then the photometrically determined 
mass ratio would be an under-estimate of the
actual value as the times of the inner eclipse
contacts will be pushed to smaller phases.

\begin{table*}
\begin{scriptsize}
\caption{Parameters obtained by direct, Roche-model BF fitting to 
the observed BFs of AW~UMa and V566~Oph. 
 \label{tab04}}
\begin{center}
\begin{tabular}{lcccc} 
\hline
Parameter                & AW~UMa        & AW~UMa          & AW~UMa           &   V566~Oph        \\
Case                     &   (1)         &    (2)          &     (3)          &    (4)            \\
\hline
$u_1$                    & 0.525         & 1.00            & 0.52             &   0.551           \\
$u_2$                    & 0.530         & 0.53            & 0.56             &   0.556           \\
$T_1$ [K]                & 6980          & 6980            & 6980             &   6540            \\
$T_2$ [K]                & 6901$^a$      & 6201(9)         & 6475(28)         &   6478$^a$        \\
$\alpha$                 & 0.00          & 0.00            & 0.30             &   0.00            \\
$T_0$ [HJD]              & 2452500.0315(1) & 2452500.0315(1)   & 2452500.0313(1)    &   2452500.2537(1) \\
$V_0$ [km~s$^{-1}$]      & $-9.31(12)$   &  $-9.43(12)$    & $-9.51(11)$      &   -37.19(18)      \\
$\Omega$                 & 1.9333(11)    & 1.9332(12)      & 1.9618(12)       &   2.2997(24)      \\
Fill-out                 & 0.256         & 0.353           & 0.304            &   0.435           \\
$K_1 + K_2$ [km~s$^{-1}$]& 316.95(28)    &  333.66(32)     & 321.65(31)       &   346.58(49)      \\
$M_1$ [$M_\odot$]        & 1.405(4)      & 1.636(5)        & 1.499(26)        &   1.466(6)        \\
$M_2$ [$M_\odot$]        & 0.136(1)      & 0.162(1)        & 0.161(3)         &   0.378(2)        \\
$q = M_2/M_1$            & 0.0969(5)     & 0.0990(4)       & 0.1079(4)        &   0.2575(11)      \\
$L_1$ [$L_\odot$]        & 5.88          & 6.53            & 5.92             &   3.65            \\
$L_1$ [$L_\odot$]        & 0.73          & 0.57            & 0.64             &   1.10            \\
$M_V$ [mag]              & 2.71          & 2.63            & 2.72             &   3.09            \\
$\sum w(O-C)^2$          & 0.038063      & 0.025208        & 0.031069         &   0.010844        \\
\hline
\end{tabular}
\end{center}
\end{scriptsize}
\flushleft{$^a$ $T_2$ recomputed from $T_1$ via the gravity darkening law.
Standard errors of parameters are given in parentheses. 
Errors of masses are determined from the 
errors of $K_1 + K_2$ and $q$; the uncertainty 
in the inclination has been neglected.} 
\end{table*}

Turning to the spectroscopic determinations of $q_{sp}$: Their 
main uncertainty is in the radial velocity semi-amplitude of the 
primary component, $K_1$, in $q_{sp}=K_1/K_2$,  
particularly when $q \rightarrow 0$. This results from 
$\delta^2_q = (1/K_2)^2 \delta^2_{K_1} + (q/K_2)^2 \delta^2_{K_2}$ 
where, for small $q$, the second term becomes irrelevant 
even when radial velocities of the secondary component are less 
accurate of the two ($\delta_{K_2} > \delta_{K_1}$).
The masses and luminosities of the components, as given in
Table~\ref{tab04}, are consistent with this prediction.

\section{Observed deviations from the Roche model}
\label{deviate}

We clearly see several deviations from the strict Roche 
model predictions in our analysis of the BFs for AW~UMa. They 
are documented in the BF residual maps, Fig.~\ref{fig05}.
The deviations are discussed mostly on the basis of our new
spectroscopic material, but -- to be sure of the consistency
in our assumptions -- we made also parallel analyzes
of the extant light curves of AW~UMa utilizing the material and
approaches as in \citet{Prb1999}. 

In what follows, we will use the term ``the primary (secondary) 
component'' to describe the appearance of
the respective stars in the broadening functions;
this would be in place of the correct, but long expressions such
as ``the primary (secondary) component signature 
as seen in the radial velocities''.

\begin{description}

\item 1. {\it The primary component is too narrow and 
looks ``triangular''}. The central parts of the primary component 
are systematically too narrow than predicted by the 
contact model. The peaked shape of the primary BF 
is the reason why it could not be fitted by the normal 
limb darkening law and required 
$u_1 \sim 1.0$ (Sections~\ref{differential} and \ref{direct}). 
A strongly differential rotation 
following the solar paradigm does not help
much in alleviating this problem (solution \#3 in 
Table \ref{tab04}) because the shape remains 
incorrect; particularly the ``kink'' at ~1/3 height (see the
next item) is then even more prominent in the fit
residuals.  

An explanation by a very strong
limb darkening does not have any good basis. 
The limb darkening effect is a strictly local 
atmospheric phenomenon; it is very hard to force it
to be larger because this would require an
increase in the local gradient of the source 
function. If anything, the source function can be easily 
made flatter (for example due to local kinetic
dissipation of horizontal motions or to magnetic 
energy deposition) so the limb darkening could be 
smaller than for normal, spherical stars rather than larger. 

An explanation by the differential rotation would 
force us to reject the Roche model which 
is based on the solid-body 
rotation law. Application of a consistent, 
differentially rotating equivalent 
of the Roche model is beyond the scope of this study, mainly 
because of the formidable theoretical obstacles 
how to treat the star shape and brightness 
distribution for such a case.
In this paper, we use a hybrid model by retaining the Roche 
model for the shape and for the brightness prediction 
while artificially imposing a non-solid body rotation law
for velocities, but we stress that such a model is 
not internally consistent.
We note that an introduction of a differential
rotation characterized by the parameter 
$\alpha \ne 0$ in Eq.~\ref{eqn01} was not fully satisfactory
and the quality of the fits was
inferior compared to assumption of a large limb darkening.
Moreover, an explanation by 
a strong differential rotation or some steady surface circulation 
patterns \citep{motl2006} would require very rapid flows
and strong velocity shear which -- in turn --
would cause shock waves 
in the photosphere, as argued by \citet{ander1983}.

The ``triangular'' shape of the primary is not seen by
us for the first time. \citet{ander1983} and later 
\citet{Rci1992a} appeared to see the same type of
deviations. Synthesized spectra of AW~UMa, including
changes of local profile over the surface, by 
\citet{ander1983} showed that the variations of the 
surface temperature over the surface of the contact 
binary certainly cannot cause the peculiarity in the
profile shape. 

\item 
2. {\it The primary component shows a wide ``base'' or
a large rotational-velocity ``pedestal''}.
There exists an additional broadening at the
base of the primary, reaching 
about 1/3 of its height above the baseline. It 
is enhanced by the pointed (triangular) shape of primary
main peak. It does not have an obvious explanation, but it may be
due to an equatorially distributed, hot matter, encompassing 
the whole system and possibly including the secondary
within its extent. 
Due to the orbital inclination of AW~UMa of 
only about 78$\degr$, an equatorial belt should 
be visible through most of the orbital revolution, although at
this point we cannot say much about its geometry.  
If the luminous matter is really there, its 
additional light in the system could explain the 
smaller than predicted (for $q=0.10$) amplitude of the
observed light curves of AW~UMa (see item \#5 below). 
However, the amount of light would have
to be large ($L_3 = 0.14$) to drive the photometric
mass ratio from $q \simeq 0.10$ to the popular 
value of the photometric mass ratio of $\approx 0.075$.

\item 
3. {\it The secondary component changes its shape with phase.}
The secondary appears to be too small during the transit 
phases as if it was not a full, Roche filling star, but 
only a much smaller core. Its two sides (limb darkened edges)
are of unequal depths during the transit 
with the negative-velocity definitely darker.
In addition, the secondary looks differently
in the two orbital quadratures: It shows a faintly marked
extension towards the primary at phases close to 0.25 while it
appears to have a double structure (as 
if this star was itself a double or contained 
a rotating ring) at phases around 0.75. 
This is best visible in a comparison
of the two orbital quadratures (Fig.~\ref{fig01}). 
In addition, the shape of the secondary appears to vary in 
time scales of perhaps weeks while the primary component 
remains remarkably stable in time. This 
variability is particularly well visible among spectra taken 
around phase 0.64 (Fig.~\ref{fig07})\footnote{It should be
noted that V566~Oph also shows a disturbance at phases close
to 0.65; see the deviations from the model in Figure~\ref{fig05}.}.

The orbital changes of the secondary component and particularly
its small dimensions when projected against the primary
may be interpreted in two ways: 
(i)~either the secondary underfills its Roche lobe or
(ii)~what we see as a secondary  is really an
accretion region around a compact core. The small
dimensions of the secondary would be consistent with
the discrepancy between $q_{sp}$ and $q_{ph}$, when
the latter is determined through the phase of the
inner eclipse contact.

\item
4. {\it The secondary component is too faint and/or
has a too low temperature.}
The secondary component is too faint in the BF's as
if the secondary was cooler than predicted. 
The required temperature to fit 
the secondary at quadratures would require 
a large temperature difference of components: For 
the assumed $T_1$ = 7,000 K, it demands 
$T_2 = 6,200 - 6,400$ K. This is incompatible
with observed light curves which show practically 
the same depth of minima. For instance
\citet{Prb1999} assuming $T_1 = 7,175$ K 
derived a similar $T_2 \sim 7,000$ K. 

The assumption of low
$T_2$ helps to fit the secondary profiles around 
quadratures, but predicts a relatively deep cut into
the profile of the primary component during its transit
in front of the primary. This is not observed and
the trough or depression during the transit 
(see Fig.~\ref{fig01}) is very shallow, as if the
temperature difference was in fact absent.

\begin{figure}
\includegraphics[width=84mm,clip=]{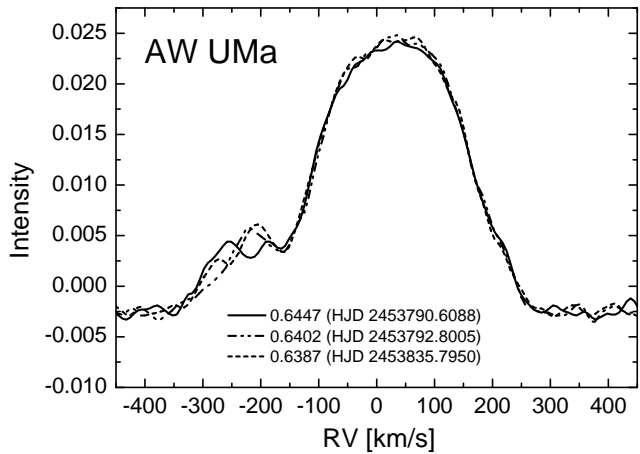}
\caption{The temporal changes of the broadening 
function around the phase 0.64 of AW~UMa. The epochs
are separated by 2 and 43 days. Note that the phase
region shown here corresponds to erratic behaviour of
residuals in Figure~\ref{fig05}.
}
\label{fig07}
\end{figure}

We suspect that this problem is not intrinsic to the
star but may be due to difficulties with 
normalization of the AW~UMa spectra prior to 
determination of broadening functions.
Similarly as for other very close binaries, the spectra
are notoriously difficult to handle in terms of 
an appropriate choice of the continuum level. This
is because spectral lines in  
strongly rotationally broadened late-type spectra 
are heavily blended forcing one to define the local
continuum relative to 
a basically arbitrary level between the lines.
The importance of this effect can be
illustrated by a simple experiment:
We convolved the template spectrum with the expected
contribution of each of the components of AW~UMa
at one of the orbital quadratures (phase 0.25).
The result is shown in Figure~\ref{fig08} where
we see directly that the real continuum is entirely
unknown. One must use a pseudo-continuum level with
an uncertainty of at least 1\% -- 2\%. 

Extensive tests of the BF technique have shown that shapes of
the BFs are entirely unaffected by the uncertainty in the
continuum level placement, but that the BF zero level may be shifted,
usually to negative values (because the continuum is
usually placed too low). The property of such simple shifts, 
without any influence on the shape, are due to the linear
characteristics of the broadening functions and are a
great advantage over the cross-correlation functions
(where zero levels are usually not accessible at all).
But the uncertainty of the zero level does influence
integrated intensities of the two stellar peaks. It 
has practically negligible influence on the 
strength of the primary component in the BF, but -- for a very
small mass ratio -- has a profound effect on the definition of
the secondary peak as small differences in the zero level
strongly influence its integrated strength. The effect is exactly
in the direction of making the secondary too faint
hence under-luminous relative to the model.

\begin{figure}
\includegraphics[width=84mm,clip=]{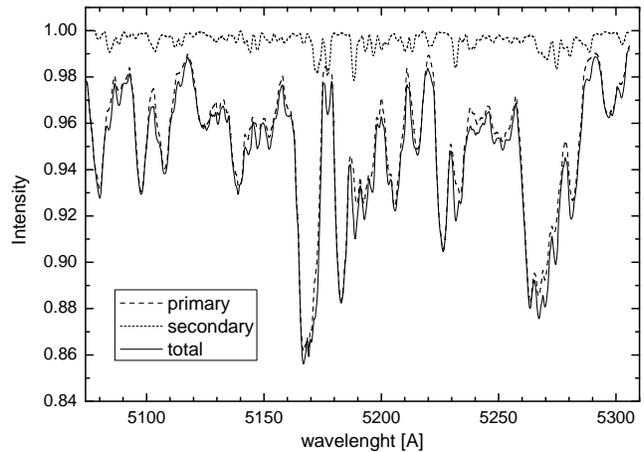}
\caption{The synthetic spectrum of the Mg~I feature in one of
the orbital quadratures (phase 0.25), obtained by separately convolving
BF contributions from each of the components of AW~UMa represented by 
rotational profiles. The individual contributions are shown by dotted 
and dashed lines; resulting spectrum by a solid line. Note that placing  
the pseudo-continuum level at, say, 0.98 (which is what one would 
probably normally do) would have a much more profound effect on the definition 
of the secondary signature in the BF than on that of the primary component.
}
\label{fig08}
\end{figure}

\begin{figure*}
\includegraphics[width=150mm,clip=]{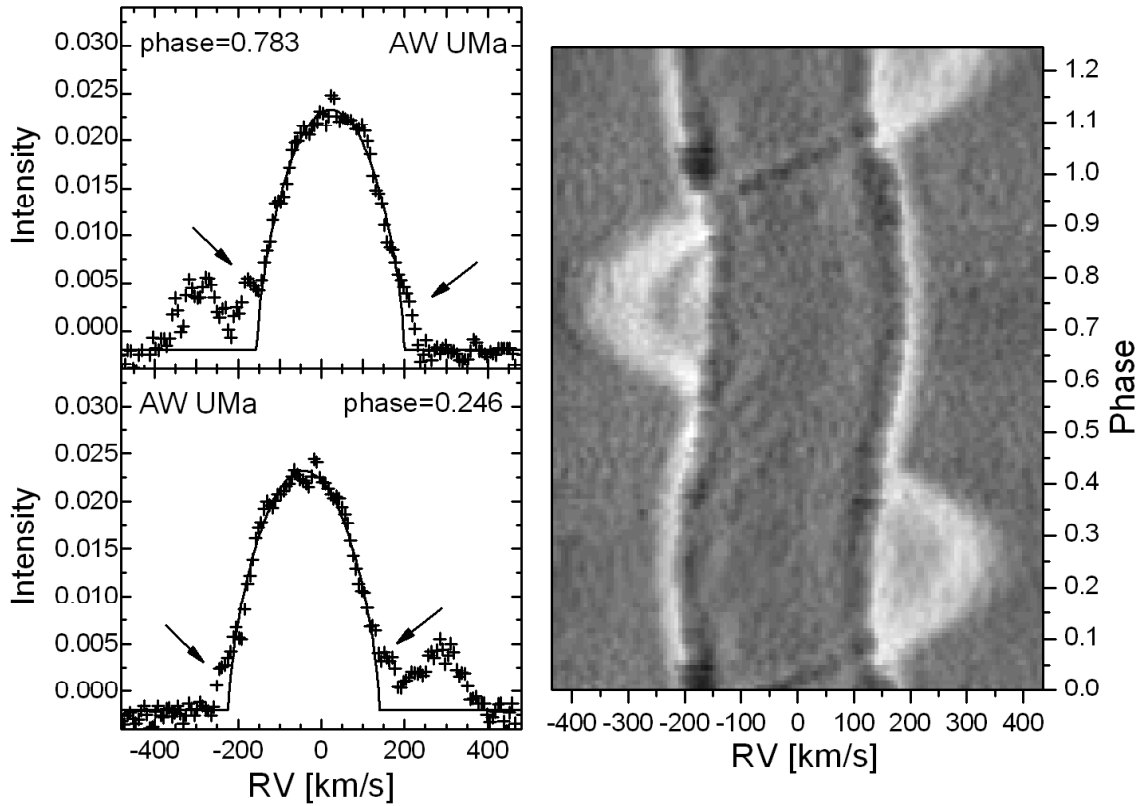}
\caption{Fits to the upper halves of the primary signature
are really excellent. Compare this figure with Figure~\ref{fig02}.
Note that the additional broadening at the base, which we
explain by the equatorial ``belt'' is now even more
prominent. The right panel shows the deviations from rotational
profile fits calculated for the primary component only (disregard
the secondary component here). The belt around the primary
shows as the light-gray edge to the profile. 
}
\label{fig09}
\end{figure*}

\item
5. {\it There is an inconsistency in the amount of light
in the system.}
It is impossible to fit the observed light curves of AW~UMa
with the parameters derived from the radial-velocity (BF)
orbital solutions. The major obstacle is the
predicted difference of the eclipse depths, 
$\delta m \approx 0.1$ mag, caused by different temperatures of 
components,  which is not observed. 
The larger mass ratio, $q \simeq 0.10$ rather than 0.075, 
forces both minima to be deeper than observed. 
All previous photometric solutions were fully consistent 
with the smaller mass ratio and
there was no contradiction between the elements. As
already mentioned, this can be alleviated by the presence of
an unaccounted third light in the system of about $L_3 \sim 0.14$. 
Except for indications of a luminous equatorial belt
(item \#2 above), we see no other possible light contribution.
A decrease of the orbital inclination to solve the
problem is out of the question because the light curve clearly
shows total eclipses.
We should stress, however, that our main conclusion concerning
the mass ratio of $q \simeq 0.10$ is based solely on the
radial velocity data and in no way is it related to the 
light amount budget evaluated from the eclipse depths.

\item 6. {\it The primary component is spotted.}
The outer (away from the secondary)
parts of the primary component reveal presence of
unexplained dark features which run diagonally in 
the phase -- radial velocity 2-D diagram in 
Figure~\ref{fig01} and can be interpreted as dark
spots on a solid-body rotating
primary. They are better visible in the picture of
the residuals in Figure~\ref{fig05}. Their presence across the
whole width of the primary component indicates that they must 
span the whole range of velocities, i.e., must be located close 
to the equator of the primary. 
They also contradict our suspicions of a 
differential rotation of the primary because they were 
apparently stabile over the whole period of one year.  
Obviously, we have no idea what is the
origin of the features that we see; if these are magnetic
spots, this would suggest a convective envelope which 
would be unexpected for the relatively early spectral type of
AW~UMa. We may note however that simple stellar-structure arguments 
suggest that the primary should have a relatively deep
convective envelope \citep{Rci1992b}.

\end{description}

It should be stressed that the BF deviations 
from the Roche model -- as seen in radial 
velocity maps -- are definitely present and
very well defined for AW~UMa, 
but are absent in V566~Oph (see for example
Fig.~\ref{fig02}). It seems that V566~Oph behaves 
fully according to the model and that it is AW~UMa 
which appears to be in some way peculiar. 
This is all strange and unexpected, mostly because of the 
excellent agreement of the AW~UMa light curve with the model.
The two binaries have very similar spectral types and orbital periods, 
and differ really only in their mass ratios,
$q = 0.10$ for AW~UMa versus $q = 0.26$ for V566~Oph. 
Is this what causes the great difference between
the systems and the strong deviations for AW~UMa?

\section{The new model for AW UMa}
\label{newmodel}

As described in the previous section, the
secondary component of AW~UMa, as seen in radial velocities,
is far from what we would expect on the basis of
the contact Lucy model. Because the secondary
is faint and dominated by the much bigger and brighter
primary companion, it is very hard to study, but its signature
is definitely very different from that expected within the 
contact model. But the well observed
primary is also showing major deviations from the contact
model. We are absolutely sure now that (1)~the span 
of the primary rotational velocities
is much less than expected for the contact model
and that (2)~there is some additional
source of continuum light at large rotational velocities
relative to the primary centre (we call it ``the belt''). 
The problem (1)~forced us to consider -- as solutions --
the unusually large limb darkening and the differential
rotation field. But both are hard to accept,
the former is not tenable on the grounds of what we know about
stellar atmospheres while the latter cannot be reconciled
with the presence of the photospheric spots strongly suggesting
that the primary does rotate as a solid body. 

\begin{figure}
\includegraphics[width=84mm,clip=]{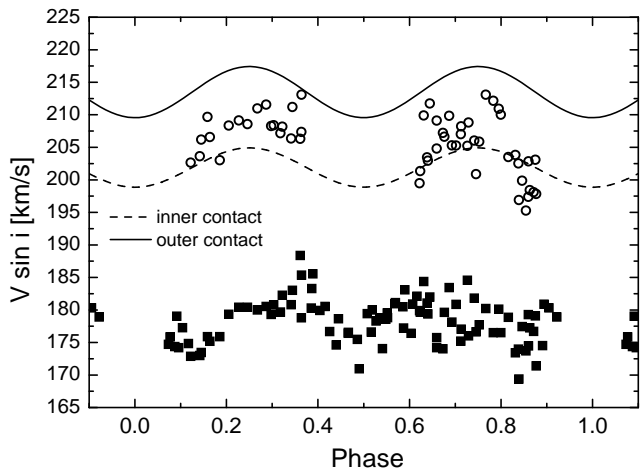}
\caption{The phase dependence of projected rotational velocity 
$V \sin i$ for the primary component. The velocities were
determined by rotational profile fits to the whole primary 
profiles, including the broadening due to the ``belt''
(open circles) and to the upper halves of the profiles (squares).
The theoretical rotational velocities of the primary component's
inner and outer critical surfaces were computed assuming
$K_1 + K_2 = 333.36$ km~s$^{-1}$ for the solution (2) in 
Table~\ref{tab04} with the fractional radii $r_{side}=0.596$ 
and $r_{back}=0.614$ with the orbital inclination angle of 78.3$\degr$.
The sine curves are simplified approximations 
of $V \sin i$ and are strictly 
valid for quadratures and conjunctions only. 
The whole profile fits were done only for phases outside
$\pm 0.12$ of the eclipses, while the upper-half fits 
excluded phases of the transit of the secondary component.
}
\label{fig10}
\end{figure}

We present here a radical and perhaps controversial 
suggestion: The primary (and probably the secondary) do not fill
their common equipotential surface and they do not form a contact
binary in the sense of Lucy's model. Our main argument is
that the whole upper half of the primary broadening function 
can be ideally fit by the simple, 
solid-body rotational profile (Figure~\ref{fig09}).
This fact, coupled with the perfectly rotationally
synchronized dark spots
on the primary, suggests that the primary is indeed a solid body
rotator with dimensions smaller than predicted by the contact model.
The rotational broadening appears to be characterized by 
by $V \sin i = 179.1 \pm  2.4$ km~s$^{-1}$ in the orbital 
quadratures and $V \sin i = 176.8 \pm 2.7$ km~s$^{-1}$ 
in the secondary minimum when the primary is seen ``from its end''. 
With the solid body rotation, the ``side'' dimension of the
primary is $ R_1(side) \simeq 1.58 R_\odot$. Thus, the 
primary's surface is close to, but some 15\%
inside its {\it inner Roche lobe\/}. Changes of
$V \sin i$ with the orbital phase (Figure~\ref{fig10})
for the primary component and for the belt indicate that
while the primary is underfilling the inner Roche lobe, the
belt may have dimensions comparable with the expected size
of the contact model; this estimate includes
the very risky assumption that the belt is also a solid body rotator.

The assumption of the detached nature of the primary component
cleanly solves the problem of the narrow, ``triangular'' shape of the
primary's BF, but at the cost of having to speculate about the ``belt''. 
We think that the belt may be an important 
source of light in the system; it is
localized in the equatorial plane, it has the surface-brightness
properties (the temperature) not much different form 
those of the stellar surfaces, it must be optically
and possibly also geometrically thick. {\it We stress that
that the belt must produce an absorption spectrum, otherwise
it would not be seen in the BF formalism\/}. There are
absolutely no indications of any emission lines or of
optical polarization \citep{Pir1975,Pir1977}; the presence of
the latter could possibly suggest electron 
scattering of the stellar spectrum.
Not much can be said about the secondary component in the 
new model, but in Section~\ref{deviate}
we listed enough indications to think that it is quite abnormal.
 
\begin{figure}
\includegraphics[width=84mm,clip=]{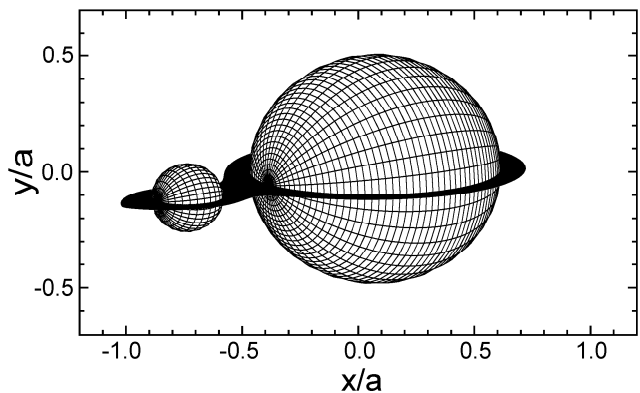}
\caption{This is how we think AW~UMA looks like. The belt in this
picture has the outer equatorial dimensions corresponding to the 
common equipotential for outer critical surface and may be
thick while both stars are within their inner Roche lobes.
}
\label{fig11}
\end{figure}

In our picture (Figure~\ref{fig11}), both
components are seen surrounded by the belt, but 
their mass centre velocities are not modified. 
Fits to the upper halves of the primary give
a well defined value for $K_1 = 30.98 \pm 0.59$
km~s$^{-1}$; this semi-amplitude is better defined
than using any of the previous measurement techniques. 
Assuming the range of the $K_2$ semi-amplitudes as 
in the columns (2) -- (4) of Table~\ref{tab03}, 
we arrive at the masses $M_1 = 1.52 - 1.59 M_\odot$ and 
$M_2 = 0.159 - 0.164 M_\odot$. These are values
practically identical to the ones determined previously,
so the final elements of AW~UMa
are not changed at all. In particular,
the mass ratio is basically the same,
$q = 0.103 - 0.104$.

The question is: Why does photometry of AW~UMa appear to be 
in such an excellent agreement with the Lucy model? And why
does the binary appear so stable if it is so complicated?
Frankly, we do not know... We prefer not to suggest
any similarities of AW~UMa to known peculiar binaries
because such analogies may constrain future options in the
interpretation. In this paper, we intended to show only
well established observational facts limiting speculations
to a minimum. We note, however, 
that a period of a strong brightness instability of AW~UMa was 
observed in 1989 -- 1990 by \citet{Derman1990}.
Thus, although the star seems to be normally stable and 
well behaving, this could be the time that the 
quiescent picture suddenly broke down.

\section{SUMMARY}
\label{sum}

Our analysis of the broadening functions for the prototype 
of extreme-mass ratio contact binaries, AW~UMa, indicates
that the generally accepted Roche model experiences 
major problems in interpretation of this system. 
We see strong indications that the system is not a
contact binary and that both stars are smaller than 
their inner Roche lobes. The interpretation is complicated by
the presence of what appears to be 
a luminous stream of the matter (optically
thick and thus giving the same stellar spectrum),
encompassing the primary and possibly 
the whole system and forming an equatorial belt 
around it. The equatorial extent of the belt
is similar to the size of the common envelope
derived in previous contact binary solutions. 

The secondary strongly changes its spectroscopic appearance 
with the orbital phase and does not appear to be a normal star 
but rather a small stellar core surrounded by
a complex disturbance of the belt. The primary is also
showing unexpected deviations from the contact model:
The ``triangular'' or 
``pointed'' shape of the broadening functions
(too narrow spectral lines) requires invoking an 
unusually strong limb darkening and/or a strongly differential 
rotation of the primary (the equatorial regions moving faster than
polar). While the first explanation is
unphysical, the second is inferior in terms of the
overall fit to the data; it is contradicted
by the presence of photospheric spots on the
primary which were stable over one year.
Our results and suggestions 
are radical ones because AW~UMa -- which
appeared to be one of the best examples of the contact
model -- would not be then a contact binary in the sense 
envisaged by Lucy and by many researchers who followed
his ideas. But we are driven to this
desperate move only after exhausting all other
alternatives.
 
A belt in the equatorial plane, encompassing both components 
is not an entirely new idea. In 1950, in his book \citet{Struve1950}
presented a picture (p.~188, Figure 28) which is very close
to what we have in mind for AW~UMa. Obviously, the new model
would question the validity of the Lucy model for at least some
binaries now considered to be in good contact. This is the price 
we appear to pay for having access to good spectral data
and to improved methods of spectral analysis. Indeed,
with only photometric information, and with the Lucy model,
the realm of contact binaries has been not only
amazingly appealing, but also much simpler.

We stress that irrespectively of assumptions used to
describe the details of the whole picture, the radial
velocities of the photocentres give 
the spectroscopic mass ratio of AW~UMa 
close to $q_{sp} \simeq 0.10$. This value is 
significantly higher than the seemingly incontestable
photometric mass ratio of $q_{ph} \simeq 0.07 - 0.08$.
By taking global averages of our results,
the masses of the components are approximately  
$M_1 = 1.5 \pm 0.15 \,M_\odot$
and $M_2 = 0.15 \pm 0.015 \,M_\odot$, 
where the uncertainties are estimated from systematic 
effects related to choices made in the analysis and
are far larger than any of the formal uncertainty  
estimates for individual methods.
Since the spectral data (in spite of several serious 
problems which we tried to stress rather than hide) 
directly give us radial velocities of the
mass centres, the spectroscopic mass ratio of 
$q_{sp} \simeq 0.10$ must be
close to the true one. The mass-ratio discrepancy,
together with direct deviations from the expected
velocity field, may be pointing to us a major 
deficiency in our assumptions about contact binaries.
At this point we cannot state whether these deficiencies are 
{\it present\/} or are merely {\it best detectable\/} 
only at the very low mass ratio of AW~UMa. In
either case, the deviations from the Roche/Lucy model 
appear to be highly significant for interpretation
of contact binary systems.

\medskip

This paper is dedicated to the memory of Bohdan Paczy\'nski,
the formidable teacher and towering intellect, but also
an understanding colleague and friend. As a student, he discovered
AW~UMa in 1963 and recognized it as a particularly important one for
understanding of contact binary stars. This star
accompanied him over 44 years of his immensely productive 
astronomical life to his last contribution which he directly
supervised \citep{BeP2007}.

We thank Stefan Mochnacki for valuable comments and suggestions and
to Kosmas Gazeas for his permission to use his DDO observations. 
The research made use of the SIMBAD database, operated at the CDS,
Strasbourg, France and accessible through the Canadian
Astronomy Data Centre, which is operated by the Herzberg Institute of
Astrophysics, National Research Council of Canada.
The stays of TP at DDO and the research of SMR
have been supported by a grant to SMR from
the Natural Sciences and Engineering Council of Canada.
This research has been supported in part by the Slovak
Academy of Sciences under a Grant No. 2/7010/7.

\label{lastpage}

\end{document}